\documentstyle[preprint,eqsecnum,aps]{revtex}

\tightenlines
\begin{document}
\draft
\title{Higher Order Effects in Electromagnetic Dissociation 
of Neutron Halo Nuclei
}

\author{S. Typel} 
\address{
National Superconducting Cyclotron Laboratory,
Michigan State University, \\
East Lansing, Michigan 48824-1321, USA
}

\author{G. Baur}
\address{
Institut f\"{u}r Kernphysik, Forschungszentrum J\"{u}lich,
52425 J\"{u}lich, Germany
}

\date{\today}
\maketitle
\begin{abstract}
We investigate higher order effects in electromagnetic excitation 
of neutron halo nuclei using a simple and realistic zero range
model for the neutron-core interaction. In the sudden (or Glauber)
approximation all orders in the target-core electromagnetic 
interaction are taken into account. Small deviations from the sudden
approximation are readily calculated. We obtain very simple 
analytical results for the next to leading order effects, which
have a simple physical interpretation. For intermediate energy
electromagnetic dissociation, higher order effects are 
generally small. 
We apply our model to Coulomb dissocation of $^{19}$C at 67~A$\cdot$MeV. 
The analytical results 
are compared to numerical results from the integration of the
time dependent Schr\"{o}dinger equation. Good agreement is obtained.
We conclude that higher order electromagnetic effects are well under
control.
\end{abstract}

\pacs{25.70.De, 25.70.Mn, 25.60.-t}

\narrowtext

\section{Introduction}
 
Electromagnetic excitation of high energy radioactive
beams is a powerful method to study electromagnetic properties 
of loosely bound neutron rich nuclei. E.g., the low lying
E1-strengths of one-neutron halo nuclei 
like ${}^{11}$Be and $^{19}$C have been studied in this way 
\cite{Ann,Nak02,Nak03}.
In a similar way, two-neutron 
halo nuclei like $^{6}$He and $^{11}$Li were studied.
Such experiments are 
usually analysed theoretically in first order electromagnetic 
perturbation theory or the equivalent photon method. In this way, 
the multipole (especially dipole) strength distribution is obtained.
Such an analysis depends on the dominance of first order excitations. 
Various methods have been developed in order to consider deviations
from first order perturbation theory with the usual multipole expansion
of the interaction. However, a consistent picture of the
importance of these approximations has still not emerged.

By ``higher order effects'' we mean only electromagnetically
induced effects on the relative momentum of the fragments. They 
can be studied in the semiclassical approximation.
In the widest sense all effects
which give rise to a deviation from the result of the
traditional semiclassical first order perturbative calculation
of Coulomb breakup can be summarized under this expression. 
In the perturbative approach higher order effects can be described
as the exchange of more than one photon 
between the target and the projectile system.
``Postacceleration'' is also a higher order effect. In a classical
picture it can be understood as a different acceleration of
the fragments in the Coulomb field of the target which will change
both the c.m.\ momentum and the relative momentum of the
particles in the final state.
In our calculations we will not treat
quantal effects like diffraction or 
contributions to the breakup from the nuclear interaction.

There are mainly two different approaches for the investigation
of higher order effects. In the semiclassical description
the projecile moves on a classical trajectory (which is usually
well justified) and experiences
a time-dependent interaction from the target. Only 
the excitation of the projectile is treated
quantally. In contrast to that, the total system
of target, projectile and fragments, respectively, 
can be described by suitable wave functions in a fully 
quantum mechanical approach. Each of these approaches has its
merits, but, at the same time, can limit the study of
certain higher order effects or make it difficult to extract
them by a comparison to a suitable first order calculation.

The breakup of the prototype of a loosely bound nucleus, the deuteron,
has for a long time been studied in the post-form DWBA theory. Later, it
has also been applied to neutron halo (core + neutron)
nuclei like $^{11}$Be. In this 
approach, the Coulomb interaction between the target and the core
is  taken into account to all orders. This is done by 
using full Coulomb wave functions in the initial and final state.
For a recent review with further 
references see \cite{TMUPROC}. 
A so called  adiabatic breakup theory
has recently been developed in \cite{adia}.
This model is related to the post-form DWBA.  
It leads to a very similar
formula, however, the physical interpretation
is somewhat different. Without entering into
the differences of the two approaches, it is clear  that
in these theories  higher order effects are  
automatically included to all orders.
It is therefore very
interesting to note that Tostevin \cite{tos} claims 
to have found substantial higher 
order effects in the Coulomb breakup of $^{19}$C \cite{Nak03}.  
He compared his 
results from the adiabatic approach to the one using semiclassical first 
order theory. Both calculations have very similar relative
energy spectra, but they differ by about 35 percent in  
absolute magnitude. 

It is the purpose of this paper to investigate higher order
effects in the electromagnetic excitation of neutron halo nuclei
by comparing lowest order and higher order approximations
{\em within} the same model. This is expected to give more reliable 
statements about the importance of these effects than the comparison
of higher order calculations in {\em one} theory with first
order calculations in {\em another} theory, where, e.g., the
finite range effects are treated in another way or the semiclassical
approximation is not applied.
We will limit ourselves
to the semiclassical description considering only the Coulomb
interaction and will not investigate nuclear induced effects.
In our approach we use a classical trajectory to describe the 
relative motion between the target and the projectile. It 
should be kept in mind that there is some
ambiguity in the definition of this trajectory. The energy loss 
should be small compared to the total kinetic energy  of the projectile
and some averaging procedure can be used. 
We assume that 
the c.m.\ of the projectile 
moves on the classical 
trajectory (straight line or Rutherford). It has been argued that
the electromagnetic interaction of the target only affects the
charged core of the projectile; therefore the c.m.\ of the core
has been used for describing the classical motion.
However, for intermediate 
energy this effect was found to be quite small in numerical calculations
in \cite{melbaye}. Actually, the result of a first order E1 calculation 
does not depend on this choice, since the dipole moment of the
system does not change. There is only a change of the 
quadrupole moment but this has small effects
since the E2 contribution to the breakup is rather small 
(see below). 
Since the c.m.\ trajectory is fixed, only higher order effects in the
relative motion of the fragments
can be handled in the semiclassical approach. 
Since the total momentum of the fragments is much larger than the
relative momentum between them, higher order effects have a much
larger effect on the relative momentum.

To some approximation the nuclear structure of neutron
halo nuclei can be described by rather simple wave functions.
Using these wave functions the reaction mechanism can be studied
in a very transparent way and analytical results are obtained. 
In a later stage more refined descriptions of the nuclei
can be introduced.
We recall some results from 
\cite{tyba} and apply the model to the electromagnetic 
breakup of $^{19}$C in comparison to more accurate descriptions.
In Ch.~2 the theoretical
framework is given; results and comparison to experimental results 
\cite{Nak03} are presented in 
Ch.~3. Conclusions are given in Ch.~4.

\section{Theoretical framework}

We follow very closely the approach of \cite{tyba}, see also \cite{babeka}.
In this straight-line semiclassical model a projectile with charge $+Ze$
impinges
on a neutron+core ($n+c$)
system with impact parameter $b$ and velocity $v$.  
The ground state wave function of the bound $n+c$ system 
is given by a simple Yukawa type wave function
\begin{equation} \label{wfbind}
\phi =\sqrt{\frac{\eta}{2\pi}} \: \frac{\exp(-\eta r)}{r}
\end{equation}
where the parameter $\eta$ is related to the binding energy $E_0$ by $E_
0= \frac{\hbar^2 \eta^2}{2 m}$ with the reduced mass 
$m=\frac{m_{n}m_{c}}{m_{n}+m_{c}}$ of the system.
The final continuum state is given by
\begin{equation}
\phi_q^{(-)} = \exp(i \vec{q}\cdot\vec{r})- \frac{1}{\eta-iq}
 \: \frac{\exp(-iqr)}{r}
\end{equation}
where the wave number $q$ is related to
the relative energy by $E_{\rm rel}=\frac{\hbar^2 q^2}{2 m}$.
With these wave functions the breakup probability can be calculated 
analytically in the sudden approximation (corresponding to the Glauber
or frozen nucleus approximation)
of semiclassical
Coulomb excitation theory including {\em all} orders in the exchange of
photons between the target and the projectile. But
the time evolution of the sytem during the excitation 
is neglected and only E1 transitions are taken into account.
The first approximation corresponds to an adiabaticity parameter $\xi$
of zero.
This quantity is the ratio of the collision time to the nuclear 
interaction time and it is given by $\xi=\frac{(E_0+E_{rel})b}{\hbar v}$.
The multipole response of the system is characterized by effective charges
$Z_{\rm eff}^{(\lambda)}=Z_c\left(\frac{m_n}{m_n+m_c}\right)^{\lambda}$.
They become very small for higher multipolarities due to the
small ratio $\frac{m_n}{m_n+m_c}$. From a perturbation expansion
of the excitation amplitude it can be shown that also the
second order E1-E1-amplitude is much larger than the first order
E2 amplitude. The ratio is given by the Coulomb (or Sommerfeld)
parameter $\frac{Z Z_{c} e^{2}}{\hbar v}$ which is much larger than one
for high charge numbers $Z$.
Therefore we can safely neglect E2 excitation in the 
following. (This is, e.g., qualitatively different for p+core systems  
like $^8$B $\rightarrow$ $^{7}$Be + p with much larger E2 effective charges.)
This can be considered as a justification of the model of \cite{tyba}. 

We expand the analytical results for the excitation probability
of \cite{tyba} for $\xi=0$ up to second  order in E1 excitation or
equivalently in the  the characteristic 
strength parameter which is given by 
\begin{equation}
 \chi=\frac{2ZZ_{\rm eff}^{(1)}e^{2}}{v b \hbar k}
\end{equation} 
where 
$k=\sqrt{\eta^2+q^2}$.
In leading order (LO), the sudden limit of the first order result
of \cite{tyba} is obtained.
Deviation for finite values of $\xi$ can be calculated according to 
\cite{tyba}. From eq.~(12) of \cite{tyba} one sees that the $\xi$
dependence of the amplitudes 
is given by $\xi K_{1}(\xi)$ or $\xi K_{0}(\xi)$ with modified 
Bessel functions,
respectively. For $\xi =0$ this 
factor (for $K_{1}$) is
1 and drops to zero exponentially for $\xi \gg 1$. 
The $\xi$-correction
in the next-to-leading order (NLO) 
goes essentially like the square of this, so we can 
only overestimate the higher order effects in the present procedure.

Instead of using the strength parameter $\chi$ we define
in the following 
the slightly different parameters 
\begin{equation}
 y= \chi \frac{k}{\eta}
\end{equation}
(independent of $q$)
and
\begin{equation}
x=\frac{q}{\eta} = \sqrt{\frac{E_{\rm rel}}{E_{0}}} \: .
\end{equation}
After angular integration over the relative momentum
between the fragments the LO breakup probability
is found to be
(see eq.~37 of \cite{tyba})
\begin{equation} \label{dpdqlo}
\frac{d P_{LO}}{d q} = \frac{16}{3\pi \eta} y^2  
\frac{x^4}{(1+x^2)^4} \: .
\end{equation}
(Note that for a correct normalization of the breakup probability the results
of \cite{tyba} have to be devided by $(2\pi)^{3}$.)
The NLO contribution is proportional to $y^{4}$ and contains 
a piece from the second order E1 amplitude and a piece from the interference 
of first and third order amplitudes. 
Again, in terms of the variables $\eta$, $x$, and 
$y$ one obtains
\begin{equation} \label{dpdqnlo}
\frac{d P_{NLO}}{d q} = \frac{16}{3\pi \eta} y^4 
 \frac{x^2(5-55x^2+28x^4)}{15 (1+x^2)^6}
 \: .
\end{equation}
The LO-expression is directly proportional to the B(E1)-strength with its
characteristic shape in the zero range model. The NLO-contribution
will introduce a change of that shape. It is weighted most in collisions
with the smallest possible impact parameters $b$ and can easily be evaluated.
For $ 0.309 < x < 1.367$ 
the NLO contribution becomes negative with the largest
reduction at a relative energy close to the binding energy.
This is essentially due 
to the interference of first and third order amplitudes. The second order
E1-E1 contribution is positive definite. From \cite{tyba} we find 
\begin{equation}
\frac{d P_{E1-E1}}{d q}=\frac{16}{3 \pi  \eta} y^4
\frac{x^2 (5+5x^2+16x^4)}{15(1+x^2)^6} \: .
\end{equation}
A reduction of the cross section at small relative energies
is only obtained if third-order contributions in the breakup
amplitude are considered,
either in a perturbative treatment (cf.\ figs.~2 - 4 in \cite{tyba2}) 
or a full dynamical calculation (cf.\ figs.~5 + 7 in \cite{EBB}).
In our analytical results we can directly see the dependence
of higher order effects on the impact parameter $b$, the projectile
velocity $v$ and the binding energy $E_{0}$ charactrized by $\eta$.
For larger impact parameters the first order E1 contribution will 
dominate more and more ($y \propto b^{-1}$). 
Perhaps, experimental accuracy will not be high enough  to see  
such a change of the shape of the breakup bump. The scaling variable y
also displays very clearly the dependence on the binding energy, 
characterized by
$\eta$, and the charge number $Z$. Since $\frac{Z_c}{m_c}$ is about 
constant for all nuclei, the breakup probabilities for heavier 
nuclei (like the r-process nuclei) are expected to be of the same 
order of magnitude as the light ones (like $^{11}$Be or $^{19}$C). 
This will be an interesting field for future RIA facilities,
where intensive beams of medium-energy neutron-rich nuclei
will become available. This is of special interest for the
r-process \cite{erice}.

The breakup cross section can be obtained by
multiplying the differential breakup probability with the
Rutherford scattering cross section
$\frac{d\sigma_{R}}{d\Omega}$ 
and the density of final states.
It is given for the LO approximation by
\begin{equation} \label{dsdo}
 \frac{d^{2}\sigma_{LO}}{dE_{\rm rel}d\Omega} = 
 \frac{d\sigma_{R}}{d\Omega} \:
 \frac{dP_{LO}}{dq} \:
 \frac{m}{\hbar^{2}q} 
\end{equation}
and similarly for the NLO approximation. 
In order to have a quick estimate of higher order effects in the total 
breakup cross section  we can integrate over the scattering angle
and the breakup relative energy 
\begin{equation}
  \sigma  =  \int dE_{\rm rel} \: \int d\Omega \:
\frac{d^{2}\sigma}{dE_{\rm rel}d\Omega}
= 2\pi \int dq \: \int\limits_{b_{\rm min}}^{b_{\rm max}} db \: b \: 
 \frac{dP_{LO}}{dq} 
\end{equation}
where we have introduced minimum and maximum impact parameters 
$b_{\rm min}$ and
$b_{\rm max}$, respectively.
We use $b_{\rm max}=\frac{\hbar v}{E_{0}+E_{\rm rel}}$ 
corresponding to a cutoff at an adiabaticity parameter of $\xi=1$.
The integration over the impact parameter $b$ is now easily
performed.
Introducing the effective strength parameter
\begin{equation}
\chi_{\rm eff} = y \eta b = \chi k b = \frac{2ZZ_{\rm eff}^{(1)}e^{2}}{\hbar v}
\end{equation}
and the minimal adiabaticity parameter
\begin{equation}
 \xi_{\rm min} = \frac{E_{0}b_{\rm min}}{\hbar v}
\end{equation}
we finally obtain
\begin{equation} \label{slo}
 \sigma_{LO} = \frac{\pi}{18} 
 \left(\frac{\chi_{\rm eff}}{\eta}\right)^{2}
 \left[ 1 - 6 \ln (4\: \xi_{\rm min})\right]
\end{equation}
and 
\begin{equation} \label{snlo}
 \sigma_{NLO} = - \frac{\pi}{18} \: b_{\rm min}^{-2}
 \left(\frac{\chi_{\rm eff}}{\eta}\right)^{4}
 \left[ \frac{23}{40}+18 \: \xi_{\rm min}^{2}\right]
 \: ,
\end{equation}
i.e., a reduction of the first order result.
The total cross sections can only be a rough guide
because of the simple treatment of the cutoff. 
Modifications due to a more precise treatment of the
$\xi$-dependence have usually to be introduced
(see below eq.~(\ref{phixi})).
However, the ratio
gives a reasonable approximation to the higher order effects.
It is a simple function depending
on the characteristic parameters of the excited system and the
experimental conditions. 

\section{Application to the Coulomb breakup of $^{19}$C}

In a recent experiment at RIKEN the breakup of $^{19}$C
into $^{18}$C and a neutron scattered on a Pb target with a beam 
energy of 67~A$\cdot$MeV was studied  and the binding energy of the
neutron was determined to be 0.53~MeV \cite{Nak03}. We apply
our model to this case since the high beam energy together with the
simple structure and  small binding
energy of the neutron is favourable for a comparison.
Another example would be $^{11}$Be where essentially the same
considerations apply.

Besides our simple analytical model we also present results
from usual first order semiclassical calculations and higher order
calculation where the full time dependent Schr\"{o}dinger equation
is solved 
for the $^{19}$C system which is perturbed by the time dependent Coulomb
field of the Pb nucleus. Here we have used the methods
described in \cite{typzfn}. 
The neutron in the bound state of
$^{19}$C was assumed to be in a $2s_{1/2}$ state
as deduced by Nakamura et al.\ \cite{Nak03}. 
The wave function
was calculated assuming a Woods-Saxon potential of radius
$r=3.3$~fm and diffuseness parameter $a=0.65$~fm. The depth is adjusted to
$V=-39.77$~MeV in order to get the experimentally extracted
binding energy of 0.53~MeV \cite{Nak03}.
As in the analytical model we use plane waves for the scattering
wave functions in the final states for $l>0$. The wave functions are
represented on a grid with exponential increasing mesh size similar
to \cite{melbaye} with 400 points up to a maximum radius of 900~fm.

In Fig.~1 we show the 
double differential cross section as a function 
of the relative energy for three scattering angles. 
We have chosen
$0.3^{\circ}$, $0.9^{\circ}$, and $2.7^{\circ}$, which corresponds
to impact parameters of $109.7$~fm, $36.6$~fm, and $12.2$~fm,
respectively. 
In order to compare the cross section in our model 
with finite-$\xi$ results of the first order semiclassical
calculation we multiply the analytical cross section given in
eq.~(\ref{dsdo}) with the shape function
\begin{equation} \label{phixi}
 \phi(\xi) = \xi^{2} \left[ K_{0}^{2}(\xi) + K_{1}^{2}(\xi) \right]
\end{equation}
of the photon spectrum and a normalization factor $N$.
The function $\phi(\xi)$
gives the correct dependence on the adiabaticity parameter 
in first order. We have $\phi(0)=1$ and it drops to zero
rapidly for $\xi > 1$.
The factor $N$ accounts for finite range
effects. The ground state wave function in the analytical
model is a $1s_{\frac{1}{2}}$ state which has a different
asymptotic normalization but the same slope as compared
to the corresponding wave function from the Woods-Saxon potential.
The results of the numerical calculation (dotted and dot-dashed lines)
agrees very well with the $\xi$-corrected cross section in the
analytical model (solid and dashed lines) for $N=2.73$. 
The slope of the wave function
is determined by the binding energy (see eq.~(\ref{wfbind}))
which is the same in both models.
There are noticeable
differences between the analytical model and the dynamical
calculation only for large relative energies and scattering angles.
The first order E2 contribution (multiplied with 1000)
is also shown in Fig.~1 (long-dashed line). 
It is at least three orders of magnitudes smaller
as compared to the first order E1 excitation cross section and can safely
be neglected. Furthermore we observe that the cross section 
decreases strongly with increasing scattering angle.
For small scattering angles results of first and higher order
calculations are almost identical. With increasing scattering angle we notice
a reduction of the cross section for small relative energies
due to higher order electromagnetic effects.

In Fig.~2 we compare the ratio of higher order (i.e., all orders in the
dynamical calculation or
LO+NLO in the analyical model, respectively) to first order
cross sections depending on the relative energy for the same
scattering angles as in Fig.~1. The solid line gives the result of
the analytical model. (Notice that the ratio is independent of $\phi(\xi)$
and $N$.) The dependence  of the ratio on the relative energy
again agrees well with
the ratio in the full semiclassical model (dotted line). 
At small relative energies there is a reduction of the
cross section (except for energies close to zero)
whereas at higher relative energies we find a small
increase.  This behaviour can be directly understood by inspecting
eqs.~(\ref{dpdqlo}) and (\ref{dpdqnlo}).
Higher order effects are largest for large scattering angles
corresponding to impact parameters close to grazing scattering.
A look at the breakup probabilities (\ref{dpdqlo}) and (\ref{dpdqnlo})
shows that higher order effects
increase essentially with $b^{-2}$. 
The discrepancy of the
two models at higher relative energy,
where the exact form of the
wave function for small radii in the range of the nuclear potential
becomes important,
is not very essential because 
the absolute cross sections are very small.
In contrast, at small relative energies, the models agree very well
since the main contribution to the matrixelements is determined
by the asymptotic form of the wave function.

Integrating the double differential cross sections over scattering
angles between $0^{\circ}$ and $3^{\circ}$ ($b_{\rm min}\approx 11$~fm)
leads to the energy
dependent cross sections in Fig.~3(a).
Again we observe that both the
first order and higher order calculations in the $\xi$-corrected 
analytical model and the full semiclassical model agree very well
in the peak of the distribution. Here we find a reduction of the
cross section of at most 10\% at small relative energies as shown
in Fig.~3(b). The spectral shape is not severely distorted.
Higher order Coulomb effects cannot explain the difference
of first order theoretical calculations and the experimental results
with respect to
the absolute magnitude and the 
shape of the
experimental data \cite{Nak03}. Although they
lead to a decrease of the cross section at small energies
(apart from the region just above threshold)
and an increase at higher energies the slope of the
theoretical results is much steeper as compared to the
experiment. In contrast, the position of the peak is well described
since it is determined
by the binding energy of the neutron in $^{19}$C.

Integrating over relative energies 
the effects of higher order are washed out and become even smaller
in the total cross section $\sigma$. We obtain 1.44~b (1.39~b) in the
first order (dynamical) semiclassical calculation and 1.49~b
(1.44~b) in the LO (NL+NLO) analytical model with $\xi$-correction,
respectively, for energies up to 3~MeV. 
Comparing to the 
experimental value of $\sigma=(1.34\pm 0.12)$~b \cite{Nak03}
one has to take into account our simple nuclear model. In reality
the ground state of ${}^{19}$C has a more complicated structure
than a $2s_{1/2}$ single particle state. Multiplying the cross sections
of our calculation with a spectroscopic factor of $0.67$ as given
by Nakamura et al.\ \cite{Nak03} the total cross section would be smaller
than in the experiment but the peak region in Fig.~3(a) would be
well described. At higher relative energies
nuclear contributions could be present in the experimental data,
increasing the total cross section again.
A possible Coulomb-nuclear interference effects could also lead to
a change of shape of the cross section.
Furthermore, the experimental data could contain contributions from
final states with an excitation of the core ${}^{18}$C.
Our results correspond to a reduction of the total cross section
by higher order effects of $3.3\%$ 
in the semiclassical model and  of $3.2\%$ in the $\xi$-corrected
analytical model. From equations (\ref{slo}) and (\ref{snlo})
we predict a 2.9\% reduction which is close to the more refined
models.
From the comparison we conclude that our simple analytical 
model with finite-$\xi$ correction 
is quite realistic in the prediction of higher order effects
and gives a reliable estimate of the reduction of the total cross section.
The smaller value of the reduction obtained in the simpler fully analytical
model can be well understood. Without taking the adiabatic suppression
correctly into account contributions to the total cross section from higher
relative energies and larger impact parameters,
where higher order effects are smaller, are not
sufficiently reduced and lead to an underestimate.
However, higher order effects in the triple differential cross section
in the peak of the excitation function are well described by the simple
analytical expressions.

Our results are in contrast to \cite{tos} where a much bigger effect
of the order of 30 to 40 percent was found by comparing different
models. It is difficult to assess how much of the reduction
is caused essentially by higher order electromagnetic effects or by
differences in these models.
In both cases, the total contribution of higher orders to the cross section
is negative.

Finally, let us make some remarks about post-acceleration.
A semiclassical model might suggest that the parallel
momentum distribution of the core is shifted towards larger values
due to an ``extra Coulomb push'' see, e.g., \cite{babeka}. However, 
this turns out to be wrong.
In the sudden approximation, the core-neutron-binding is negligible 
and also on its 
way towards the target, the core alone (and {\em not} the bound
core-neutron system)
feels the Coulomb deceleration. Formally, this is easily seen:
In the sudden approximation, the momentum transfer points exactly to the
direction 
perpendicular to the trajectory, 
the excitation amplitude depends only on 
$\vec{q} \cdot \Delta \vec{p}$ (cf.\ \cite{tyba}).
This is symmetric with respect to the 
plane perpendicular to the beam direction. Corrections  of this
simple result due to small  values of $\xi$ were studied in \cite{tyba}.
They were found to depend only on the phase shift of the neutron
s-wave. This phase shift is given in the analytical model by 
$\delta_{0} = - \arctan \frac{q}{\eta}$.
It is a rather delicate quantal interference effect and even
has the opposite sign to what one would have thought ``intuitively''. 
Large values of $\xi$ correspond to large values $b$ where the
strength parameter is small. Therefore,  higher order effects
are not so important. 
Indeed, in Ref. \cite{promptor} no effects of
post-acceleration were found for the $^{11}$Be system. 

\section{Conclusions}

We have studied the basic example for the Coulomb dissociation
of a neutron halo nucleus. From the simple zero-range wave function
of a loosely bound system it becomes directly obvious that the 
low lying E1 strength is an immediate consequence of the halo
structure.
It is probably the most beautiful manifestation of the halo
nature. Higher order effects can be described by analytical
formulas. This allows a very transparent discussion of the
effects. Our results can be easily applied to all neutron halo
Coulomb dissociation experiments. They are a useful guide for
the much more elaborate numerical solutions of the time-dependent
Schr\"{o}dinger equation.
Our  simple considerations are corroborated by these
more sophisticated approaches. We conclude that
higher order electromagnetic effects
are not a significant problem  in medium-energy Coulomb dissociation
experiments and can be kept under control.

\acknowledgments

The authors would like to thank R. Shyam for useful discussions and
B. A. Brown and P. G. Hansen 
for helpful remarks on the manuscript. We are grateful to
T. Nakamura for providing us with the experimental data.
Support for this work was provided from US National Science
Foundation grant numbers PHY-0070911 and PHY-9528844.

\begin{figure}
\caption{Double differential cross section for the Coulomb dissociation
of 67~MeV/u $^{19}$C scattered on $^{208}$Pb as a function of the relative
energy for three scattering angles.
Analytical model with finite
$\xi$-correction: LO-calculation (solid line), LO+NLO-calculation
(dashed line); semiclassical calculation: E1 first order (dotted line),
E1 dynamical (dot-dashed line), E2 first order multiplied by 1000
(long-dashed line).}
\label{fig1}
\end{figure}

\begin{figure}
\caption{Ratio of higher order to first order
double differential cross sections for the Coulomb dissociation
of 67~MeV/u $^{19}$C scattered on $^{208}$Pb as a function of the relative
energy for three scattering angles. Analytical model with finite
$\xi$-correction (solid line) and semiclassical calculation
(dotted line).}
\label{fig2}
\end{figure}

\begin{figure}
\caption{(a) Differential cross sections integrated over
scattering angle from $0^{\circ}$ to $3^{\circ}$ 
for the Coulomb dissociation
of 67~MeV/u $^{19}$C scattered on $^{208}$Pb as a function of the relative
energy.
Analytical model with finite
$\xi$-correction: LO-calculation (solid line), LO+NLO-calculation
(dashed line); semiclassical calculation: E1 first order (dotted line),
E1 dynamical (dot-dashed line); experimental data from \protect\cite{Nak03}.
(b) Ratio of higher order to first order differential 
cross sections for the $\xi$-corrected analytical model (solid line) and the 
dynamical model (dotted line).}
\label{fig3}
\end{figure}

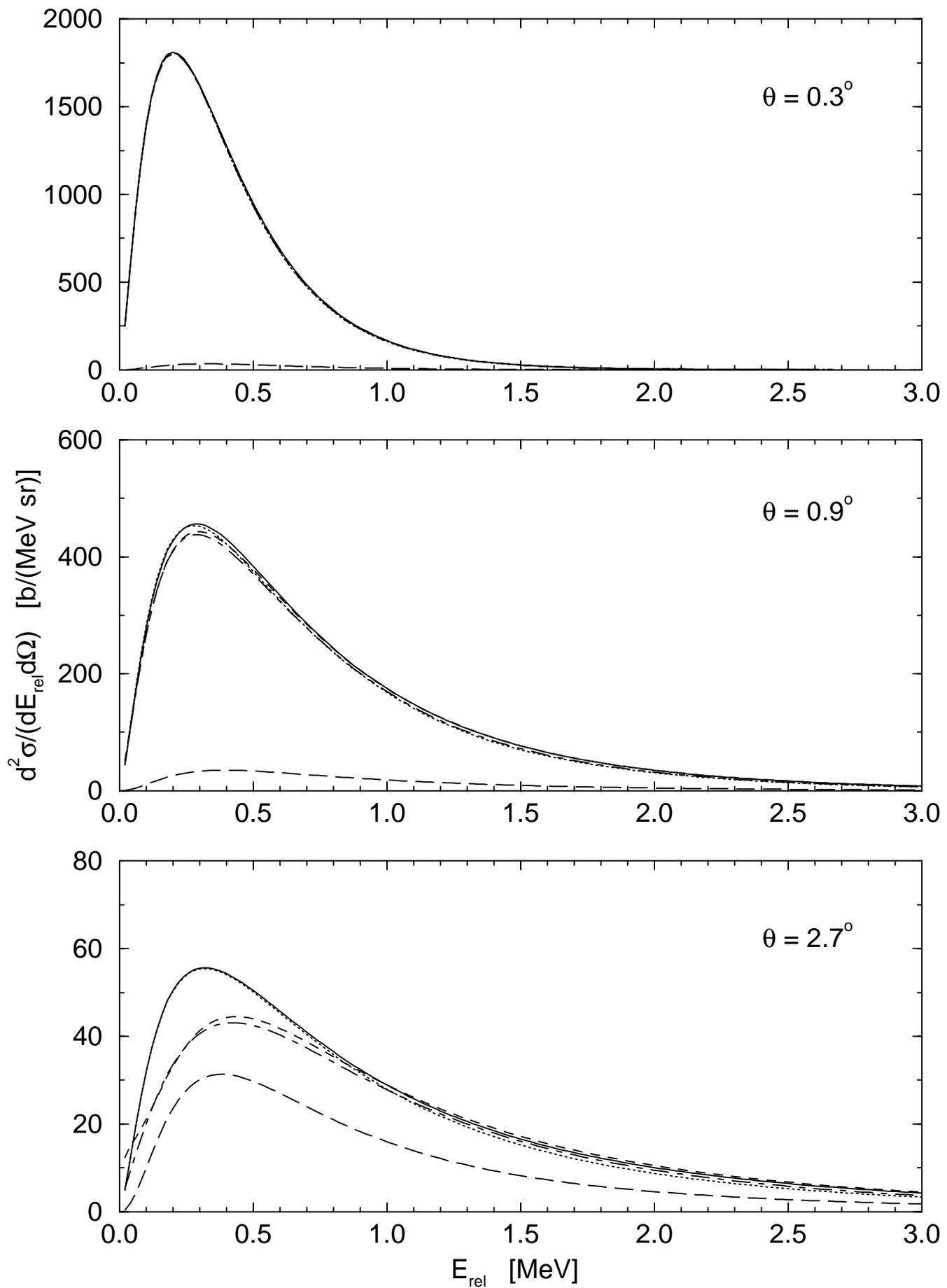
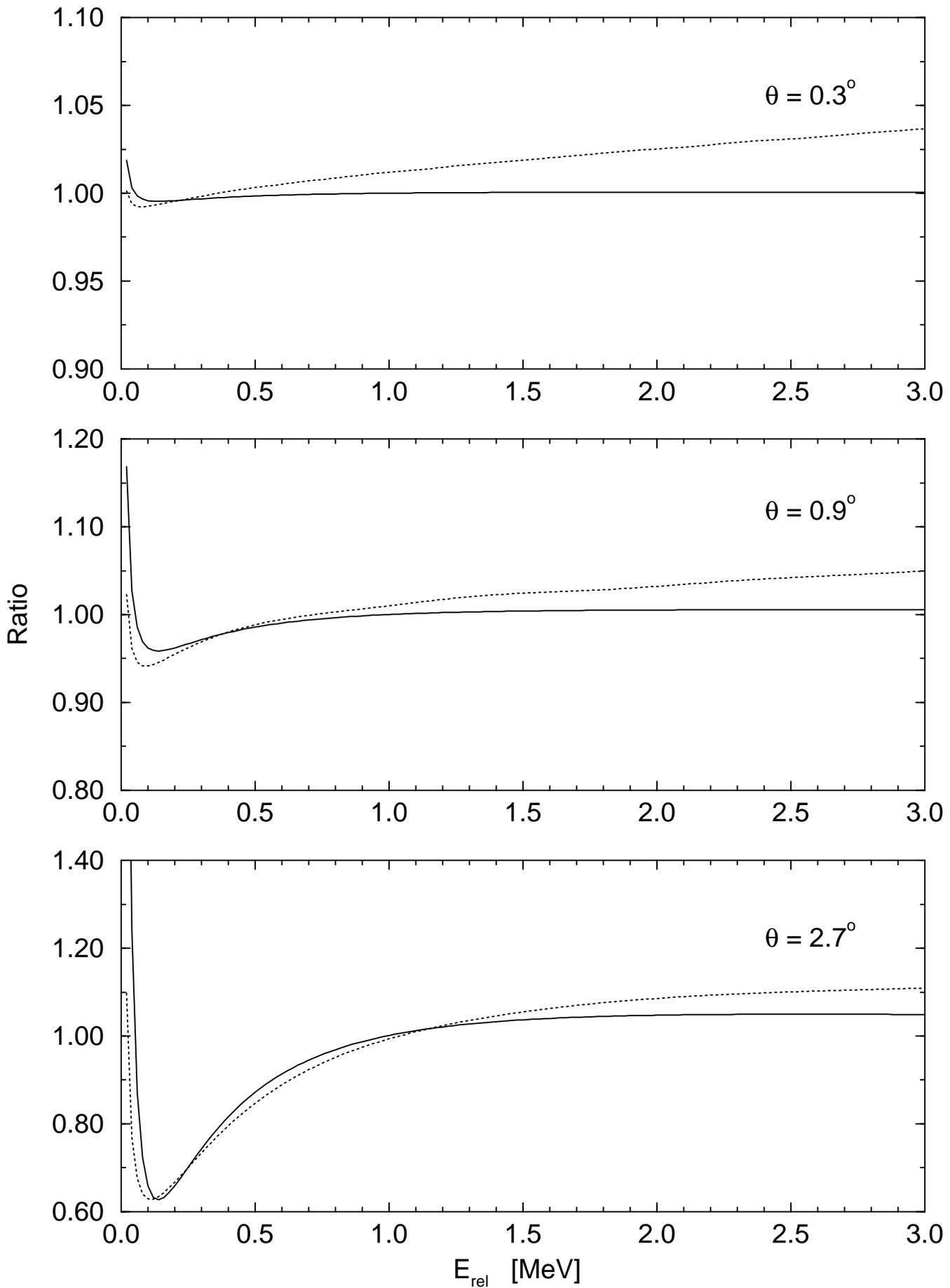
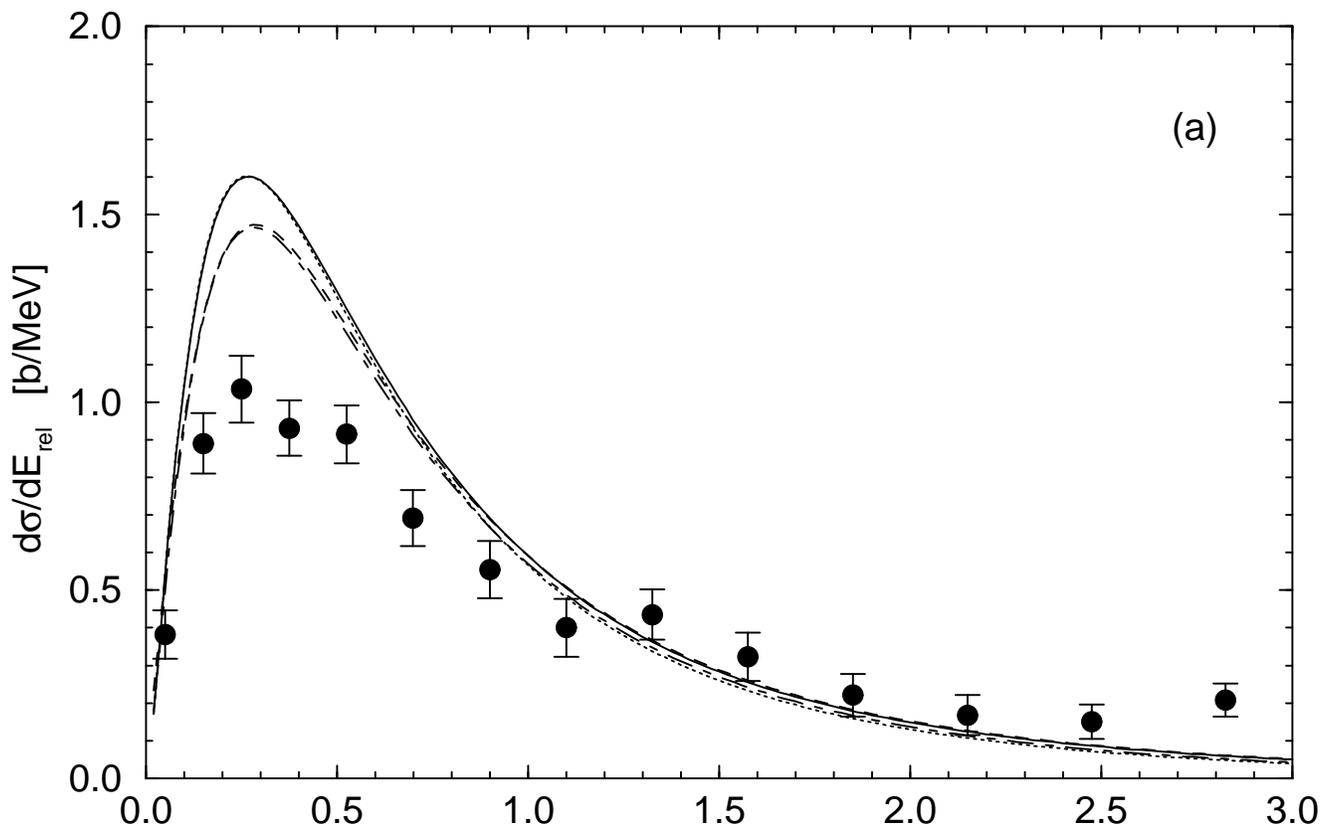
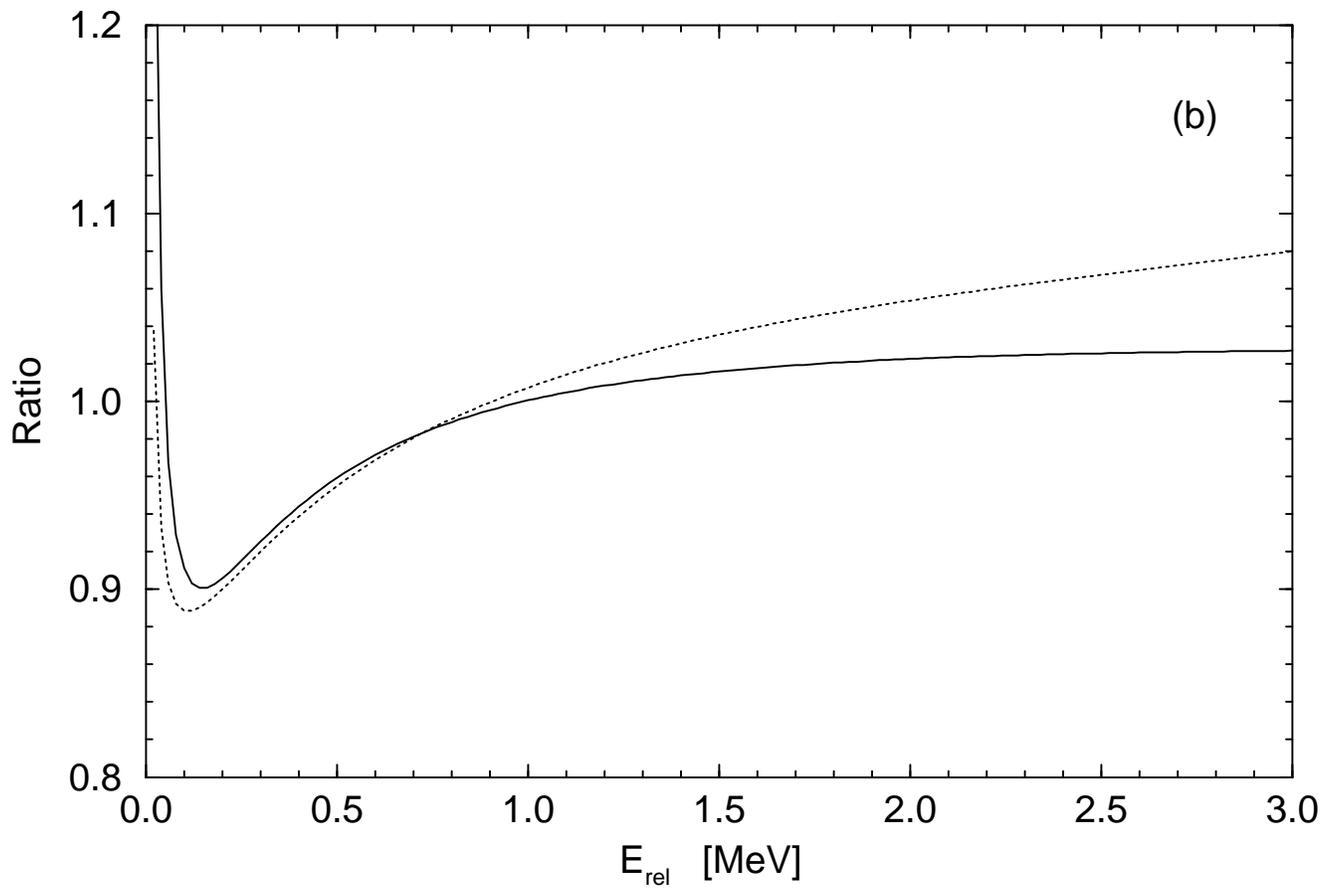
\end{document}